\documentclass[aps, showkeys,showpacs,nofootinbib,floatfix]{revtex4}

\usepackage{amssymb}
\usepackage{amsmath}
\usepackage{graphicx}
\usepackage{hyperref}
\usepackage{graphicx}
\begin{document}

\title{Spectrums of Black Hole in de Sitter Spacetime with Highly Damped Quasinormal Modes: High Overtone Case}
\author{Molin Liu$^{1}$}
\thanks{Corresponding author\\E-mail address: mlliu@xynu.edu.cn}
\author{Xuehui Hu$^{1}$}
\author{Junwang Lu$^{2}$}
%\author{Benhai Yu$^{1}$}
\author{Jianbo Lu$^{2}$}
\affiliation{$^{1}$College of Physics and Electronic Engineering,
Xinyang Normal University, Xinyang, 464000, P. R. China\\
$^{2}$Department of Physics, Liaoning Normal University, Dalian, 116029, P. R. China}

\begin{abstract}
Motivated by recent physical interpretation on quasinormal modes presented by Maggiore\cite{Maggiore}, the adiabatic quantity method given by Kunstatter\cite{Kunstatter} is used to calculate the spectrums of a non-extremal Schwarzschild de Sitter black hole in this paper, as well as electrically charged case. According to highly damped Konoplya and Zhidenko's numerical observational results for high overtone modes\cite{Konoplya}, we found that the asymptotic non-flat spacetime structure leads two interesting facts as followings:
(i) near inner event horizon, the area and entropy spectrums, which are given by $A_{en} = 8 n_1 \pi \hbar$, $S_{en} = 2\pi n_1\hbar$, are equally spaced accurately.
(ii) However, near outer cosmological horizon the spectrums, which are in the form of $A_{cn} = 16 n_2 \pi \hbar - \sqrt{\frac{48\pi}{\Lambda}A_{cn} - 3 A_{cn}^2}$, $S_{cn} = 4 \pi n_2 \hbar - \sqrt{\frac{3\pi}{\Lambda}A_{cn} - \frac{3}{16} A_{cn}^2}$, are not markedly equidistant. Finally, we also discuss the electrically charged case and find the black holes in de Sitter spacetime have similar quantization behavior no matter with or without charge.
\end{abstract}

\pacs{04.70.Bw, 04.70.Dy, 04.62.+v}

\keywords{area and entropy spectrums; highly damped quasinormal modes; de Sitter spacetime}

\maketitle
%\maketitle  IS IGNORED %%%%%%%%%%%
\section{Introduction}
Issue regarding black hole (BH) quantization problem has been intensively and extensively focused on. As early as the 1970s, Bekenstein conjectured that if the horizon
area is an adiabatic invariant, the horizon
area should have a discrete and equally spaced spectrum \cite{Bekenstein1,Bekenstein2,Bekenstein3},
\begin{equation}\label{o1quantumarea}
    A_n = \epsilon l_p^2 n, \ \ \ \ n = 0, 1, 2, \cdots,
\end{equation}
where $\epsilon$ is a dimensionless constant and $l_p$ is the Planck length. The gravitational units of $c = G = 1$ and $l_p^2 = G\hbar/c^3 $ are adopted. It shows that any classical adiabatic invariant should be corresponded to a quantum entity with discrete spectrum according to the Ehrenfest principle. After that, many people have obtained the spectrums for various dynamical modes with various parameter $\epsilon$ including $\epsilon = 8\pi$ and $\epsilon =  4 \ln n$ \cite{oldwork1,oldwork2,oldwork3,oldwork4,oldwork5,oldwork6}.

Since the quantization of BH is put forth, people find that the quasinormal modes (QNMs) has a very important place in Loop Quantum Gravity. Hod firstly observed that the spacing
parameter $\epsilon$ could be determined by the high overtones quasinormal (QN) frequencies \cite{Hod1,Hod2}. Subsequently,
Kunstatter suggested that the system adiabatic
invariant should have a form of $E/\Delta \omega$ in which $E$ is system
energy and $\Delta \omega$ is vibrational frequency \cite{Kunstatter}. It is interesting that
if we choose the real part of QN frequencies $\omega_R$ and the ADM mass $M$ as the vibrational frequency $\Delta \omega$ and the system energy $E$, Hod's result $\epsilon = 4 \hbar \ln 3$ is obtained naturally through Kunstatter method.

Recently, Maggiore considers above situations and proposes that
the QN frequencies should be treated as the proper frequency of the
equivalent harmonic oscillator \cite{Maggiore}. The change of the
mass of BH $\Delta M$ is given by $\Delta M = \hbar \Delta \omega$
where $\Delta \omega $ is identified with the change between
adjacent physical frequencies ($m \longrightarrow m -1$) where $m$ is overtones number. According
to the harmonic oscillator theory, the actual frequency is $\omega(E) = \sqrt{|\omega_I|^2 + |\omega_R|^2}$, rather than original $\omega_R$. For the highly damping limit $|\omega_I|\gg |\omega_R|$, one has a proper frequency $\omega = |\omega_I|$.

Although the usual QNMs are important for observational aspect of gravitational waves phenomena, the high overtones QNMs are typically ignored because these relevant modes damped infinitely fast do not significantly contribute to the gravitational wave signal. However, since Hod opens a new era connecting BH quantization and QNMs, the highly damped QNMs ($m \gg 1$) has been attracted more attentions. Motl subsequently presents these asymptotic QNMs in Ref.\cite{Motl1} by continued fractions methods assuming the
gauge group of the theory to be SO(3) \cite{Leaver}. However, when $m
\longrightarrow \infty$ the
numerical method becomes more hard to solve. Hence, Motl-Neitzke developed the monodromy method by matching several poles in the plane which can calculate the highly damped QN frequencies for
any static spacetime in principle \cite{Motl2}.

On the other hands of this paper, more astronomical observations \cite{expandcosmology1,expandcosmology2} demonstrate us a nonzero cosmological constant for the expanded universe whose
topology structure presents more like a de Sitter case (for the review see Refs.\cite{cosmologicalreview1,cosmologicalreview2}). Schwarzschild de Sitter (SdS) spacetime becomes more and more important in the gravity theory. Recently, more attentions are paid on the near extremal
SdS BH in which the BH event horizon and the cosmological horizon
are close to each other
\cite{Cardosoextreme,Molina,Maassen,Yoshida,Liwenbo}. In 2003 \cite{Cardosoextreme},
Cardoso-Lemos obtained their QN
frequencies for various radiative fields including scalar
perturbation, electromagnetic perturbation and gravitational
perturbation by using P$\ddot{o}$shl-Teller potential with very
large imaginary part \cite{Ferrari}. Then this result
\cite{Cardosoextreme} has been used to quantize the area and entropy
of the near extremal SdS BH \cite{Setare,Liwenbo}. In early Setare's work \cite{Setare}, the
real part of QN frequencies is considered to be a fundamental
vibrational frequency for a BH of energy. The area spectrum of near
extremal case above is $A_b = 24 \pi n \hbar \sqrt{V_0/\kappa^2_b -
1/4}$ where $V_0$ is dependent on the form of perturbations: namely $V_0 =
l (l + 1) \kappa^2_b$ for scalar or electromagnetic perturbations, $V_0 = (l + 2) (l - 1) \kappa^2_b$ for gravitational perturbation. In the recent work \cite{Liwenbo}, Li-Xu-Lu (LXL) have used Maggiore's physical
interpretation of quasinormal modes and restudied this near extremal case. They found the area spectrum is $A_n = 8\pi n
\hbar$ and the entropy spectrum is $S_n = 2 \pi n$ which are independent of the form of perturbations. Comparing the near extremal case \cite{Setare,Liwenbo} with other systems \cite{Maggiore,Wei,Kunstatter}, one will find it is similar to the single horizon BH because the inner horizon and outer horizon are degenerate into a extremal single horizon case, similar to Schwarzschild case. So under the same Maggiore's interpretation, the result of Ref.\cite{Liwenbo} is in agreement with the results of Schwarzschild BH \cite{Wei}. However, as far as we known there is no work relevant to the quantization for the more general SdS spacetime with non extremal cosmological constant. So based on above situations including Maggiore's interpretation and highly damped QNMs of SdS BH, we use Kunstatter method to calculate the spectrums of area and entropy for SdS BH.
\section{Highly damped quasinormal modes of Schwarzschild de Sitter black hole}
The static spherically symmetric metric of SdS BH is given by
\begin{equation}\label{2metric}
    d s^2 = -f(r) d t^2 + f^{-1}(r) d r^2 + r^2 d \Omega_2^2,
\end{equation}
where lapse function is
\begin{equation}\label{2metricfunction}
    f(r) = 1 - \frac{2 M}{r} - \frac{r^2}{L_{ds}^2},
\end{equation}
$M$ is BH mass and $L_{ds}^2 = 3/\Lambda$ is de Sitter curvature radius. The spacetime above is bounded by two horizons, namely, inner event horizon $r_e$ and outer cosmological horizon $r_c$, which are listed as followings,
\begin{equation}
\left\{
\begin{array}{c}
r_{c} = \frac{2}{\sqrt{\Lambda}}\cos\eta ,\\
r_{e} = \frac{2}{\sqrt{\Lambda}}\cos(120^\circ-\eta),\\
\end{array}
\right.\label{re-rc}
\end{equation}
where $\eta = 1/3 \arccos(-3M\sqrt{\Lambda})$ with $30^\circ
\leq\eta\leq 60^\circ$. The real solutions are accepted
only if $\Lambda$ satisfies $\Lambda M^2\leq 1/9$ \cite{Liu1}.
The lapse function (\ref{2metricfunction}), hereby, could be rewritten as,
\begin{equation}\label{23metricfunc}
    f(r) = \frac{1}{L_{ds}^2 r} (r - r_e)(r_c - r)(r - r_o),
\end{equation}
where the negative parameter $r_o = - (r_e + r_c)$ has no physical meaning.

Matching Eqs.(\ref{2metricfunction}) and (\ref{23metricfunc}), we can get the relationships of BH mass $M$ and de Sitter curvature radius $L_{ds}$ as
\begin{eqnarray}
  L_{ds}^2 &=& r_e^2 + r_e r_c + r_c^2, \label{ldss} \\
  2 M L_{ds}^2 &=& r_e r_c (r_e + r_c). \label{MLds}
\end{eqnarray}
According to the variations of Eqs.(\ref{ldss}) and (\ref{MLds}), we have the formulas with respect to $\delta r_c$, $\delta r_e$ and $\delta M$ as,
\begin{equation}\label{deltamhorizons}
    (r_c - r_o) (r_e -r_c) \delta r_c = (r_e - r_o) (r_c - r_e) \delta r_e = 2 L_{ds}^2 \delta M.
\end{equation}

Then, we need introduce the surface gravity $\kappa_{i}$ and the tortoise coordinate $x$ in the following words. According to the definition of surface gravity
\begin{equation}
\label{surgravdefin} \kappa_{i}=\frac{1}{2}\left|\frac{df}{dr}\right|_{r=r_i},
\end{equation}
one can easily obtain its explicit forms
\begin{eqnarray}
 \label{sur1} \kappa_{e} &=& \frac{(r_{c}-r_{e})(r_{e}-r_{o})}{6r_{e}}\Lambda, \\
 \label{sufacecc} \kappa_{c} &=& \frac{(r_{c}-r_{e})(r_{c}-r_{o})}{6r_{c}}\Lambda,
  %K_{o}= \frac{(r_{o}-r_{e})(r_{c}-r_{o})}{6r_{o}}\Lambda.
\end{eqnarray}
where $\kappa_{e}$ and $\kappa_{c}$ are the surface gravity at $r_e$ and $r_c$, respectively.
Submitting Eqs.(\ref{sur1}) and (\ref{sufacecc}) into Eq.(\ref{deltamhorizons}), we can find
the relationship between $\delta r_c$ and $\delta r_e$
\begin{equation}\label{adddeltarecre}
    \delta M =  \kappa_{e} r_e \delta r_e = -  \kappa_{c} r_c \delta r_c.
\end{equation}
Here ``$-$" denotes the inverse relation between $r_e$ and $r_c$, which indicates when BH mass $M$ increases ($\delta M > 0$), $r_e$ increases ($\delta r_e > 0$), but $r_c$ decreases ($\delta r_c < 0$). Here, it is convenient to use Eqs.(\ref{deltamhorizons}) and (\ref{adddeltarecre}) to calculate the spectrums of area and entropy.

Through the definition of tortoise coordinate shown by
\begin{equation}
x=\frac{1}{2M}\int\frac{dr}{f(r)},\label{tortoise}
\end{equation}
we can get
\begin{equation}
 x = \frac{1}{2M}\bigg{[}\frac{1}{2\kappa_{e}}\ln\left(\frac{r}{r_{e}}-1\right)-\frac{1}{2\kappa_{c}}\ln\left(1-\frac{r}{r_{c}}\right) + \frac{1}{2\kappa_{o}}\ln\left(1-\frac{r}{r_{o}}\right)\bigg{]}.\label{tor-grav-sf}
\end{equation}

By using above coordinate Eq.(\ref{tor-grav-sf}) the perturbation field equation of
$\Phi = 1/r e^{i\omega t}\chi (r) Y_{lm}(\theta, \phi)$
could be reduced to a form of Schr\"{o}dinger wave-like equation as
\begin{equation}\label{KKr}
    - \frac{d^2 \chi}{d x^2} + V_l (r) = \omega^2 \chi,
\end{equation}
where $\chi$ is the radial component and $V_{L} (r)$ is potential given by
\begin{equation}\label{potentialpt}
    V_{L} (r) = \left(1 - \frac{2 M}{r} - \frac{r^2}{L_{ds}^2}\right) \left(\frac{2M}{r^3} - \frac{2}{L_{ds}^2} + \frac{l(l + 1)}{r^2}\right),
\end{equation}
which dominates the evolution of field in SdS spacetime. One of outstanding characteristics of this potential is that when $x$ approaches two horizons, the potential $V_L(r)$
goes exponentially to zero.

Based on the potential (\ref{KKr}), various QNMs could be solved out by adding some special boundary conditions \cite{cardosophdthesis}. In the early studies on the aspect of perturbation focus mainly on the dynamics of radiative field such as
two-dimensional toy models \cite{twotoy1,twotoy2,twotoy3}, radiative three time
evolution epochs \cite{Brady1}, radiative tails \cite{Brady2},
degenerate special Nariai BH \cite{Bousso,Nariai1,Nariai2} and so on. In order to calculate the spectrums of area and entropy, the highly damped QNMs must be found from them. Fortunately, there is an available case could be used for the computation of these spectrums, namely the high overtone case which is observed not only by Konoplya-Zhidenko's numerical observations in Ref.\cite{Konoplya}, but also by Cardoso-Nat$\acute{a}$rio-Schiappa's theoretical analysis in Ref.\cite{Cardoso111} as well as Choudhury-Padmanabhan's work in Ref.\cite{Choudhury}.

The accurate high overtones numerical quasinormal modes have been obtained by Konoplya-Zhidenko (KZ) in Ref.\cite{Konoplya} by using Leaver approach \cite{Leaver} and Nollert's \cite{Nollert1} techniques. One interesting phenomenon from these results is that the average value of spacing over sufficiently large number of modes equals event horizon surface gravity $\kappa_e$, which could be expressed mathematically as
\begin{equation}\label{singlearea}
    \sum^{m = N_1}_{m = N_2} \frac{Im(\omega_{m + 1}) - Im(\omega_m)}{N_2 - N_1} \approx \kappa_e, \ (m\ is\ very\ large).
\end{equation}
This result is in agreement with the analytical results presented by Cardoso-Nat$\acute{a}$rio-Schiappa in Ref.\cite{Cardoso111}. Simultaneously, Choudhury-Padmanabhan \cite{Choudhury} also study the corresponding analytic analysis of level spacing of the QNM frequencies and provide a derivation of the imaginary parts of frequencies for the SdS spacetime by calculating
the scattering amplitude in the first Born approximation. It is found that the numerical observations and theoretical analysis all strongly indicate the spacing of high overtone's imaginary part is proportional to the surface gravity of inner event horizon, rather than the surface gravity of the cosmological
horizon in the view of actual physics. Certainly, the situation ``scattering
off the cosmological horizon"  can happen in mathematics, which can be found in the part of Appendix C in Ref.\cite{Choudhury}. Here, we do not consider it.

\section{area and entropy spectrums for Non-extremal Schwarzschild de Sitter black hole}
For a given system with energy $E$, Kunstatter proposes the adiabatic invariant quantity $I$ should have a integral of vibrational frequency $\omega (E)$ as \cite{Kunstatter}
\begin{equation}\label{33action}
    I = \int \frac{d E}{\Delta \omega},
\end{equation}
where we adopt $E = M$ according to the first law of BH thermodynamics. Apparently, how to choose the frequency $\omega$ becomes the key to calculate the spectrums. In Hod's early work \cite{Hod1,Hod2}, the transition frequencies equal the classical oscillation frequencies for large
quantum numbers according to Bohr's correspondence principle. The highly damped BH oscillations
frequencies have a very large imaginary part for $m \longrightarrow
+ \infty$. It indicates the effective relaxation time for BH
returning to static state is arbitrarily small. The degraded rapidly QNMs
should be treated as one type of quantum transition in which the
transition frequency is the real part of $\omega$ in a natural
way.

However, according to Maggiore's interpretation of QNMs \cite{Maggiore}, the Black hole perturbations
should be looked as the damped harmonic vibration. As a result, the
real frequencies of the equivalent damped harmonic oscillators are
$\omega_0 = \sqrt{\omega_R^2 + \omega_I^2}$ rather than $\omega_R$, where $\omega_R$ is the real part and $\omega_I$ is the
image part of $\omega$. For the long lived modes $\omega_I
\longrightarrow 0$, the real frequencies become $\omega_0 =
\omega_R$. Contrary to long lived case, the frequencies become $\omega_0 = |\omega_I|$ for the highly excited modes $\omega_I \gg \omega_R$. Then, we consider the transition from $m$ mode to $m - 1$ mode for a non-extremal cosmological constant. Thus according to the numerical observations (\ref{singlearea}) the transition frequency is $\Delta \omega = \kappa_e$.

Submitting the surface gravity $\kappa_e$ (\ref{sur1}) into Eq.(\ref{33action}), we can get the Kunstatter's adiabatical invariant $I$ shown as
\begin{equation}\label{11kunstattical}
 I = \frac{1}{\Lambda} \sin \left(\frac{\pi}{6} + 2\eta\right).
\end{equation}
According to the positive quantization numbers in Bohr-Sommerfeld condition,
the adiabatic invariant quantity $I$ has to be adopted the absolute value of integral (\ref{33action}) to compute the area and entropy spectrums through Kunstatter method \cite{Kunstatter}. If we consider the expresses of inner horizon $r_e$ and outer horizon $r_c$, i.e. Eq.(\ref{re-rc}), adiabatically invariant $I$ is presented in two forms as,
\begin{eqnarray}
  I(r_e) &=& \frac{1}{2} r_e^2, \label{11Ire}\\
  I(r_c) &=& \frac{1}{4} r_c^2 + \frac{r_c}{4} \sqrt{\frac{12}{\Lambda} - 3r_c^2}.\label{11Irc}
\end{eqnarray}
The constants appeared in above formulas (\ref{11Ire}) and (\ref{11Irc}) could be assimilated into integral constants.

Hence, if the Bohr-Sommerfeld (BS) quantization condition is adopted, Eqs.(\ref{11Ire}) and (\ref{11Irc}) are rewritten as
\begin{eqnarray}
 n_1 \hbar &=&  I(r_e) = \frac{1}{2} r_e^2,\\
 n_2 \hbar &=&  I(r_c) = \frac{1}{4\Lambda} \left[\Lambda r_c^2 + r_c \sqrt{\Lambda\left(12-3\Lambda r_c^2\right)}\right],
\end{eqnarray}
where $n_1 = 1, 2, \cdots$ and $n_2 = 1, 2, \cdots$. By using the area formulas: $A_e = 4\pi r_e^2$ and $A_c = 4\pi r_c^2$, the quantum areas of inner event horizon $A_{en}$ and outer cosmological horizon $A_{cn}$ are written as
\begin{eqnarray}
 A_{en} &=& 8 n_1 \pi \hbar,\label{22aen}\\
 A_{cn} &=& 16 n_2 \pi \hbar - \sqrt{\frac{48\pi}{\Lambda} A_{cn} - 3 A^2_{cn}},\label{22acn}
\end{eqnarray}
which lead to the spacing of area spectrums shown by
\begin{eqnarray}
  \Delta A_e &=& A_{e n+1} - A_{e n} = 8 \pi \hbar,\\
  \Delta A_c &=& A_{c n+1} - A_{c n} = 16 \pi \hbar + g (A_{c n,n+1}),
\end{eqnarray}
where $g (A_{c n,n+1})$ is a correction term to the spacing shown by
\begin{equation}
g (A_{c n,n+1}) = \sqrt{\frac{48\pi}{\Lambda}A_{c n} - 3 A^2_{c n}} - \sqrt{\frac{48\pi}{\Lambda}A_{c n + 1} - 3 A^2_{c n + 1}}.\label{ggacnn1}
\end{equation}
The entropies $S_{en}$ and $S_{cn}$ of horizons $r_e$ and $r_c$ are listed as,
\begin{eqnarray}
\label{senn}  S_{en} &=& 2\pi n_1 \hbar, \\
\label{scnn}  S_{cn} &=& 4\pi n_2 \hbar - \sqrt{\frac{3\pi}{\Lambda} A_{cn} - \frac{3}{16}A_{cn}^2}.
\end{eqnarray}
It is interesting that the spectrums of area and entropy are equally spaced near inner event horizon. However, the spectrums are not equidistant near outer cosmological horizon.

Then, we have a look at the near extremal degenerate situation where $r_c$ is very close to $r_e$. Mathematically, this limit case is corresponding to the specific parameters: $M^2 \Lambda = 1/9$ and $\eta = 60^\circ$ which lead the limits of $r_e$ and $r_c$,
\begin{equation}
  r_e \sim r_c \sim \frac{1}{\sqrt{\Lambda}}. \label{22rerc}
\end{equation}
Submitting Eq.(\ref{22rerc}) into above Eqs.(\ref{11Ire}) and (\ref{11Irc}), we can obtain
\begin{equation}
  I(r_e) \sim I(r_c) \sim \frac{1}{2\Lambda}.\label{2Ire1}
\end{equation}
The quantum areas of horizons $r_e$ and $r_c$ are given by
\begin{equation}\label{22quanareas}
    A_{en} = A_{cn} = 8\pi n \hbar.
\end{equation}
The entropies also reduce to
\begin{equation}\label{22quanentropss}
    S_{en} = S_{cn} = 2\pi n \hbar.
\end{equation}
It is clear that Eqs.(\ref{22quanareas}) and (\ref{22quanentropss}) are exactly accordance with LXL's results appeared in Ref.\cite{Liwenbo}.

So according to the accurate high overtones numerical quasinormal modes observations (\ref{singlearea}), the area and entropy spectrums ($A_{en}$, $S_{en}$), ($A_{cn}$, $S_{cn}$) are obtained at event horizon $r_e$ and cosmological horizon $r_c$, respectively. For the non-extremal SdS BH, ($A_{en}$, $S_{en}$) are equally spaced near $r_e$ and ($A_{cn}$, $S_{cn}$) are not marked equidistant. However, for the near-extremal SdS BH, the horizons $r_e$ and $r_c$ coincide and the area and entropy spectrums ($A_{en}$, $S_{en}$) and ($A_{cn}$, $S_{cn}$) are reduced to the same forms with equally spacing, namely Eqs.(\ref{22quanareas}) and (\ref{22quanentropss}). In the SdS BH case, we find the behavior of QNMs frequency decides the results of area and entropy spectrums. For the very high overtones numerical quasinormal modes, the spacing of frequency is proportional to the surface gravity of event horizon $\kappa_e$, and not $\kappa_c$. Hence, this fact determinate the spacing of area and entropy spectrums is constant at event horizon $r_e$, and is changed at cosmological horizon $r_c$.

At the last part of this section, we need to analyse the not equidistant area spectrums of cosmological horizon $A_{cn}$. According to the based expressions Eq.(\ref{22acn}), we can get
\begin{equation}\label{adddletaaccc}
 \Delta A_c = A_{c n+1} - A_{c n} = \frac{\pi\sqrt{3}}{24} \sqrt{- 4 n_2^2 \hbar^2 +  \frac{4 n_2 \hbar}{\Lambda}+ \frac{3}{\Lambda^2} } \left(\sqrt{1 + \frac{-4 (2n_2 + 1)\hbar^2 + 4\hbar/\Lambda}{-4n_2^2 \hbar^2 + 4n_2\hbar/\Lambda + 3/\Lambda^2}} - 1\right).
\end{equation}
In order to observe the variation of spacing $|\Delta A_c|$ with $n_2$ and $\Lambda$ more clearly, we plot two figures below, one is  $|\Delta A_c|$ vs $n_2$ in Fig.\ref{fg1} and another is $|\Delta A_c|$ vs $\Lambda$ in Fig.\ref{fg2}. In the Fig.\ref{fg1}, we fix the cosmological constant and draw the spacing of cosmological horizon area $\Delta A_c$ versus quantum number $n_2$, which illustrates that $\Delta A_c$ decreases with increasing $n_2$. At the same time, in the Fig.\ref{fg2}, we fix the quantum number $n_2$ and draw the spacing of cosmological horizon area $\Delta A_c$ versus cosmological constant $\Lambda$, which illustrates that $\Delta A_c$ also decreases with increasing $\Lambda$.
\begin{figure}
  \includegraphics[width=4.5 in]{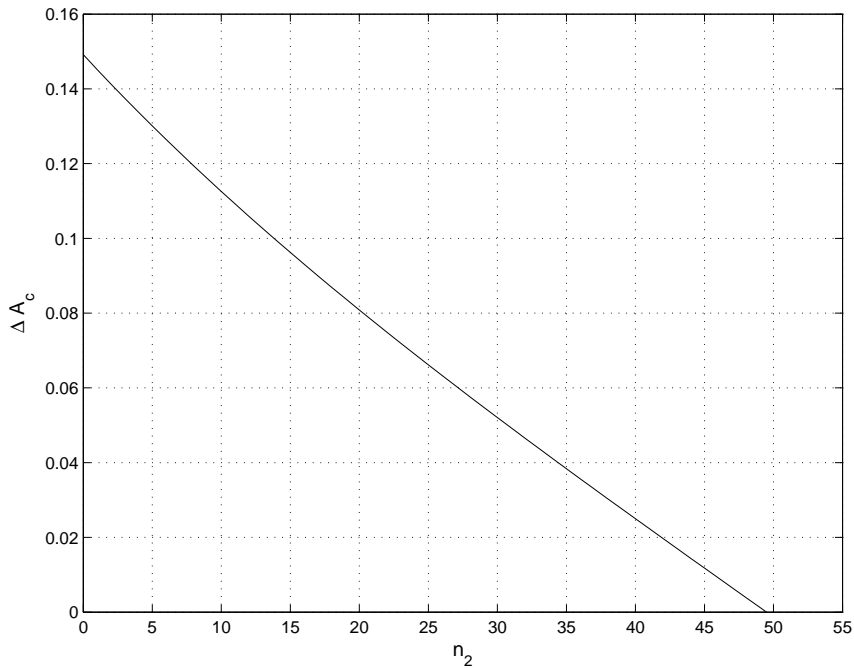}\\
  \caption{spacing of cosmological horizon area $\Delta A_c$ versus $n_2$ with $\Lambda = 0.01$ where we put $\hbar$ equal to unity.}\label{fg1}
\end{figure}
\begin{figure}
  \includegraphics[width=4.5 in]{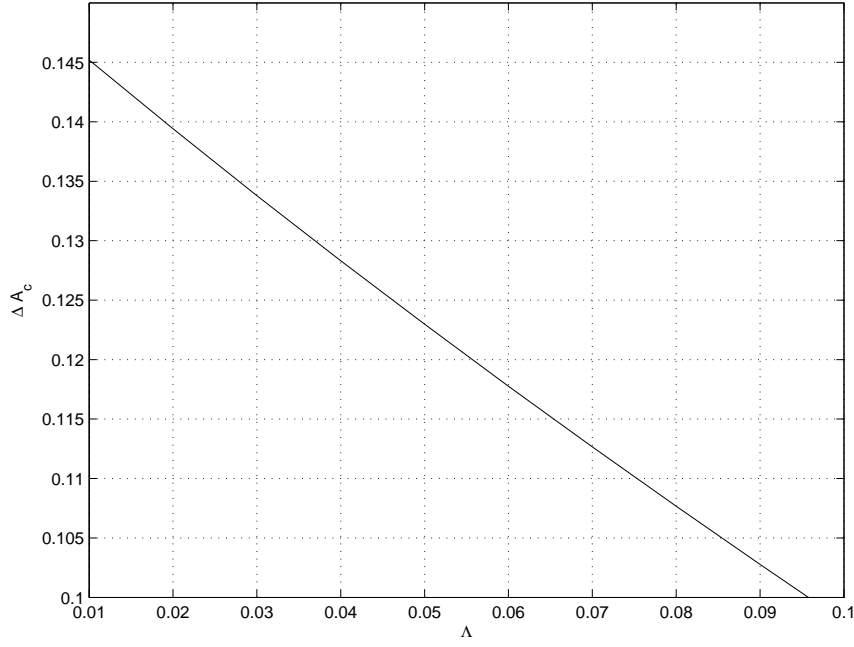}\\
  \caption{spacing of cosmological horizon area $\Delta A_c$ versus $\Lambda$ with $n = 1$ where we put $\hbar$ equal to unity.}\label{fg2}
\end{figure}

At the end, it is well known that in the de Sitter spacetime another important type black hole contained electrically charge is the Reissner Nordstr$\ddot{o}$m de Sitter black hole. One natural question is whether the behavior of above quantization of SdS BH is still valid for the electrically charged case. So we study this subject in the next section.
\section{More General Setting: Electrically Charged Case}
For a more rounded consideration, we expand the situation from no electrically charged SdS case to electrically charged case in this section. Based on the QNMs analysis \cite{Molina00,Molina,wangbwb}, we investigate the quantization of Reissner Nordstr$\ddot{o}$m de Sitter black hole (RNdS BH). In the relevant QNMs analysis, it is found that there is a great deal of observational evidence for the existence fact that the influence of charge on field propagation is mild, such as the constant tail for scalar perturbations with $l = 0$ \cite{Brady1,Brady2,Molina00}, the range of variation of the quasinormal modes \cite{Molina00}, the exponential coefficients of exponential late time tail decay \cite{Molina00} and so on.

For RNdS BH the function $f(r)$ is given by
\begin{equation}\label{addRNdS}
    f(r) = 1 - \frac{2M}{r} -\frac{r^2}{L^2_{ds}} + \frac{q^2}{r^2}.
\end{equation}
Unlike the SdS case, there are three horizons: Cauchy horizon $r_-$, event horizon $r_+$ and cosmological horizon $r_c$ in this spacetime with relationship $r_- < r_+ < r_c$. It should noticed that the results Eqs.(\ref{ldss}) and (\ref{MLds}) only hold for SdS case, for the RNdS case the Eqs.(\ref{ldss}) and (\ref{MLds}) are replaced by
\begin{equation}\label{addfrr}
    f(r) = \frac{1}{L^2_{ds}r^2} (r - r_+)(r_c - r)(r - r_-)(r - r_o),
\end{equation}
where $r_o = -(r_+ + r_- + r_c)$ with no actual physical meaning.

According to the definition of surface gravity Eq.(\ref{surgravdefin}), we can get the corresponding surface gravities $\kappa_+$, $\kappa_c$ to the horizons $r_+$ and $r_c$ shown by,
\begin{eqnarray}
\label{kkke}  \kappa_+ &=& \frac{(r_c - r_+)(r_+ - r_-)(r_+ - r_o)}{2 L^2_{ds}r_+^2}, \\
\label{kkkc}  \kappa_c &=& \frac{(r_c - r_+)(r_c - r_-)(r_c - r_o)}{2 L^2_{ds}r_c^2}.
\end{eqnarray}

Matching Eqs.(\ref{addfrr}) and (\ref{addRNdS}), the relationships of $M$, $L_{ds}$ and $q$ are obtained as,
\begin{eqnarray}
\label{addmlq1}  L^2_{ds} &=& (r_+^2 + r_+ r_c + r_c^2) + r_- (r_+ + r_c + r_-), \\
\label{addmlq2}  2M L^2_{ds} &=& (r_+ + r_c) \left[r_- (r_+ + r_- + r_c) + r_+ r_c\right],\\
\label{addmlq3}  q^2 L^2_{ds} &=& r_+ r_c r_-(r_+ + r_c + r_-).
\end{eqnarray}
It is found that if the charge vanish $q \rightarrow 0$, we can get $r_- (r_+ + r_c + r_-) \rightarrow 0$ according to Eq.(\ref{addmlq3}). Hence, the relationships above reduce to the case of SdS, namely the Eqs.(\ref{ldss}) and (\ref{MLds}). Implementing the variation of Eqs.(\ref{addmlq1}) and (\ref{addmlq3}), we can get
\begin{equation}\label{adddeltamrec}
    \left(r_c - r_+\right) \left[\left(2 r_+ + r_c\right) - \frac{q^2 L_{ds}^2}{r_+^2 r_c}\right] \delta r_+ = \left(r_+ - r_c\right) \left[\left(r_+ + 2 r_c\right) - \frac{q^2 L_{ds}^2}{r_+ r_c^2}\right] \delta r_c = 2 L_{ds}^2 \delta M.
\end{equation}
Then submitting the Eq.(\ref{addmlq3}) into Eq.(\ref{adddeltamrec}), the relationship of $\delta r_+$, $\delta r_c$, $\delta M$ are reduced to,
\begin{equation}\label{add123457774}
    \frac{1}{r_+} (r_c - r_+) (r_+ - r_-) (r_+ - r_o) \delta r_+ = - \frac{1}{r_c} (r_+ - r_c) (r_c - r_-) (r_c - r_o) \delta r_c = 2 L_{ds}^2 \delta M.
\end{equation}
Submitting Eqs.(\ref{kkke}) and (\ref{kkkc}) into Eq.(\ref{add123457774}), we can get a simple and important formulas as
\begin{equation}\label{addhuajianhoudeltarm}
    \delta M = \kappa_+ r_+ \delta r_+ = - \kappa_c r_c \delta r_c,
\end{equation}
where ``$-$" means that when $r_e$ increases $r_c$ will be decreased. It is noticed that the relationship above is obtained without any extremal assumption. Comparing with that of no-charge case it is easy to find this similar relationship about horizons and black hole mass could be also obtained in the SdS black hole case. In another words, the charge of black hole does not affect the quantization progress of black hole under de Sitter space-time. In spite of that we still need to finish the quantization of black hole contained charge in de Sitter space-time.

Then, we need to find the quasinormal frequencies which could be used calculate the area spectrums. Just like the words said by Ref.\cite{Molina}, it is usually difficult to calculate the analytic expressions for the quasinormal frequencies, except in particular situations.
Here, we must face one difficulty of that there are not clear analytic expressions about RNdS black hole quasinormal frequencies, except the near extreme limit case \cite{Molina,Molina00}.
Unlike the SdS case, there are neither purely electromagnetic nor gravitational modes. Hence, four mixed electromagnetic and gravitational fields, two of them named polar fields, $Z_1^+$ and $Z_2^+$, which impart no rotation to the black hole. And the last two are called axial fields, $Z_1^-$ and $Z_2^-$. The deduction can be found in Refs.\cite{Mellor,Molina00,Molina} and references therein.

The frequencies of quasinormal modes associated with the P\"{o}schl-Teller potential are shown by \cite{Molina00}
\begin{equation}\label{addqnmsfrequency}
    \omega_n = \sqrt{V_0 - \frac{1}{4}\kappa_+^2} - i \kappa_+ \left(n + \frac{1}{2}\right).
\end{equation}
The constant $V_0$ is denoted by $V_{0}^{sc}$ for scalar perturbation $\Phi$, or ${V_{0}^{1-},V_{0}^{2-}}$ for axial perturbations $Z_{1,2}^-$, or ${V_{0}^{1 +},V_{0}^{2 +}}$ for polar perturbations $Z_{1,2}^+$. For the scalar field, $V_0^{sc}$ is given by,
\begin{equation}\label{addspoten}
 V_0^{sc} = \frac{l(l+1)(r_c - r_+) \kappa_+}{2 r_+^2}.
\end{equation}
For the two axial potentials, $V_{0}^{1-}$ and $V_{0}^{2-}$ are given by,
\begin{eqnarray}
  V_{0}^{1-} &=&  \frac{(r_c - r_+) \kappa_+}{2 r_+^4} \left[l(l + 1) r_+^2 + 4 q^2 - r_+ S_1\right],\\
  V_{0}^{2-} &=&  \frac{(r_c - r_+) \kappa_+}{2 r_+^4} \left[l(l + 1) r_+^2 + 4 q^2 - r_+ S_2\right].
\end{eqnarray}
where
\begin{equation}\label{adds1s2}
    S_1 = 3 M - \sqrt{9 M^2 + 4(l + 2)(l - 1)q^2},\ \  S_2 = 3 M + \sqrt{9 M^2 + 4(l + 2)(l - 1)q^2}.
\end{equation}
For the two polar potentials, $V_{0}^{1+}$ and $V_{0}^{2+}$ are given by,
\begin{eqnarray}
  V_{0}^{1+} &=&  \frac{(r_c - r_+) \kappa_+}{2 r_+^4} \left[\frac{\left(2 c r_+^2 + 3 M r_+ + r_+ \sqrt{9 M^2 + 8 c q^2}\right)\left(c r_+ + M + \frac{2\Lambda r_+^3}{3}\right)}{c r_+ + 3 M - \frac{2 q^2}{r_+}}\right] + C,\\
  V_{0}^{2+} &=&  \frac{(r_c - r_+) \kappa_+}{2 r_+^4} \left[\frac{\left(2 c r_+^2 + 3 M r_+ - r_+ \sqrt{9 M^2 + 8 c q^2}\right)\left(c r_+ + M + \frac{2\Lambda r_+^3}{3}\right)}{c r_+ + 3 M - \frac{2 q^2}{r_+}}\right] + C,
\end{eqnarray}
where $C = 2 M r_+ - 2 q^2 - 2\Lambda r_+^4/3$ and $c = (l + 2)(l - 1)/2$.

Then in the last words of this part we use the analytic QNMs frequencies (\ref{addqnmsfrequency}) to derive the area spectrums of Reissner-Nordstr$\ddot{o}$m-de Sitter black hole. For the highly excited black hole, the proper frequency adopts $\omega = |\omega_I|$ with very large $n$. Hence, according to the former QNMs frequency $\omega_n$ (\ref{addqnmsfrequency}), we can know that when the system transit from $n$ to $n - 1$ the absorbed energy is
\begin{equation}\label{addadbsor}
    \delta M = \hbar \omega_I (n - 1) - \hbar \omega_I (n) = \hbar \kappa_+,
\end{equation}
which also can be treated as the transition frequency.

Submitting above Eq.(\ref{addadbsor}) into Eq.(\ref{33action}), we can get the adiabatic invariant in the form shown by,
\begin{equation}\label{addiii}
    I = \int \frac{d E}{\Delta\omega} = \frac{M}{\kappa_+}.
\end{equation}
Based on the Bohr-Sommerfeld condition, Eq.(\ref{addiii}) leads directly the equally spaced mass spectrum as
\begin{equation}\label{addmassspec}
    M = n \hbar \kappa_+.
\end{equation}
The areas of horizons $r_+$ and $r_c$ are given by
\begin{equation}\label{addarearerc}
    A_+ = 4\pi r_+^2, \ \ A_c = 4\pi r_c^2.
\end{equation}
The corresponding variations are given by
\begin{equation}\label{addareavariation}
    \delta A_+ = 8\pi r_+ \delta r_+, \ \ \delta A_c = 8\pi r_c \delta r_c.
\end{equation}
By using the former Eqs.(\ref{addhuajianhoudeltarm}), the above variations $\delta A_+$ and $\delta A_c$ is given by,
\begin{equation}\label{addare12312341}
    \delta A_+ = 8\pi \hbar, \ \ \delta A_c = \mathcal{F}(M, q, L_{ds}) 8\pi \hbar,
\end{equation}
where $\mathcal{F}(M, q, L_{ds}) = \kappa_+/\kappa_c$. Clearly the area spectra of event horizon $r_+$ is equally spaced. Then we have a look at the cosmological horizon area $A_c$, for the general case the ratio $\kappa_+/\kappa_c$ is not constant, the spaced of $A_c$ is not equally. However, for the near extremal RNdS black hole the three hypes of extremal horizons coincide, namely $r_- \sim r_+ \sim r_c$, we can get $\mathcal{F}(M, q, L_{ds}) = 1$. And $\delta A_+ = \delta A_c = 8\pi \hbar$. It is observed that the influence of the no trivial electric charge is mild. The variation of area or entropy spectrums with the charge is not very large. In another words, the charge doses having no significant effect on the quantization of black hole in de Sitter background. This type behavior of quantization is agreement with the near extremal SdS black hole case \cite{Liwenbo}. In fact, if the formula (\ref{singlearea}) can be expanded to the all high overtones no matter with charge or without charge for the black hole in de Sitter background, the spectrums of area or entropy near event horizon are equally spaced accurately, but they are not markedly equidistant at cosmological horizon. However, for the near extremal black hole in which the horizons coincide each other, all area spectrums are equally spaced.

\section{Conclusion}
In this paper, we have calculated the spectrums of non-extremal SdS BH by using Kunstatter method under Maggiore's interpretation of QNMs. We summarize what has
been achieved.

1. The QNMs plays a important role in the quest for a quantum gravity theory. The quantization of single horizon spacetimes has been attracted many people focused on including the Schwarzschild BH \cite{Hod1,Hod2,Kunstatter,Maggiore,Wei}, non-rotating or rotating BTZ BH
\cite{Wei,Kwon}, higher dimensional BH \cite{Kunstatter,Wei}. However, few works address the nontrivial multi-horizons
spacetime in non-asymptotically flat spacetime. Even for the near extremal SdS BH \cite{Liwenbo}, the close inner and outer
horizons situation leads to a usual processing method like single horizon case.
It is highly necessary to investigate the area and entropy spectrums for non-asymptotically flat SdS BH with double horizons. Hence, we perform our calculation through Konoplys-Zhidenko's result \cite{Konoplya}.

2. The spectrums of area and entropy are obtained by using Konoplys and Zhidenko's QNMs \cite{Konoplya}. The sum formula Eq.(\ref{singlearea}) reveals the statistic treatment on the average high overtone number $m$. For the gravitational perturbations case, the spacing between nearby overtones $Im(\omega_{m + 1}) - Im(\omega_{m})$ shows peculiar periodic dependence on $m$ (illustrating in Fig. 4 of Ref.\cite{Konoplya} where $m$ should be changed to $n$). But, the average value of these spacing over very large number of modes equals to the surface gravity at cosmological horizon $\kappa_e$. Hence, in the statistical viewing, the spacing of nearby overtones Eq.(\ref{singlearea}) is effective when overtone number $m$ is sufficiently large. For the electromagnetic perturbations, the spacing of $Im(\omega_{m})$ damps to a equidistant spectrum for the very high $m$, i.e. $Im(\omega_{m}) = - \kappa_e (m + 1/2)$. Then, by using high overtones results Eq.(\ref{singlearea}) above, we naturally obtain the area spectrums $A_{en}$ (\ref{22aen}) and $A_{cn}$ (\ref{22acn}) near event horizon $r_e$ and cosmological horizon $r_c$, respectively. Through the area spectrum formulas proposed by Kunstatter's \cite{Kunstatter}, we find only the spectrum of inner event horizon is equally spaced.

3. In the last part of our calculation, we expand the quantization from the no charged black hole (SdS BH) to the charged black hole (RNdS BH). After our carefully calculating, we find one important balance relationship between the black hole mass and horizons, namely $\delta M = \kappa_+ r_+ \delta r_+= -\kappa_c r_c \delta r_c$, which also can be deduced from SdS BH case. Because it is very difficult to obtain the analytic expressions for the no-extremal RNdS BH, except for in particular situation, we choose the frequencies given by Ref.\cite{Molina00}, which also is one type of near extreme case, to calculate the area spectrums according to the relevant QNMs analysis. In fact if the formulas (\ref{singlearea}) about the QNMs could be expanded to RNdS BH, the quantization of area spectrums indeed have the same behavior. However, this type verification is beyond this paper range.

\acknowledgments
This work is supported by the National Natural Science Foundation of China under Grant Nos.11005088 and 11147150.

\end{document}